\documentstyle[12pt]{article}
\newcommand{\es}{\\[2mm]}

\newcommand{\journal}[4]{{\em #1~}#2\,(19#3)\,#4;}

\newcommand{\pr}{\journal {Phys. Rev.}}

\newcommand{\jmp}{\journal {J. Math. Phys.}}

\newcommand{\cmp}{\journal {Comm. Math. Phys.}}

\newcommand{\np}{\journal {Nucl. Phys.}}
\newcommand{\pl}{\journal {Phys. Lett.}}

\newcommand{\nc}{\journal {Nuovo Cim.}}

\makeatletter
\@addtoreset{equation}{section}
\makeatother

\def\Lp{\displaystyle{\biggl(}}
\def\Rp{\displaystyle{\biggr)}}
\def\LP{\displaystyle{\Biggl(}}
 \def\wti{\widetilde}
\def\RP{\displaystyle{\Biggr)}}
\newcommand{\lp}{\left(}\newcommand{\rp}{\right)}
\newcommand{\lc}{\left[}\newcommand{\rc}{\right]}

\renewcommand{\d}{\delta}

\renewcommand{\o}{\omega} \renewcommand{\O}{\Omega}

\newcommand{\dT}{\wti d}

\renewcommand{\AA}{{\cal A}}

\newcommand{\FF}{{\cal F}}
\newcommand{\GG}{{\cal G}}

\newcommand{\JJ}{{\cal J}}

\newcommand{\NN}{{\cal N}}

\newcommand{\PP}{{\cal P}}

\newcommand{\VV}{{\cal V}}

\newcommand{\complex}{{\kern .1em {\raise .47ex
\hbox {$\scriptscriptstyle |$}}
    \kern -.4em {\rm C}}}
\newcommand{\real}{{{\rm I} \kern -.19em {\rm R}}}
\newcommand{\rational}{{\kern .1em {\raise .47ex
\hbox{$\scripscriptstyle |$}}
    \kern -.35em {\rm Q}}}
\renewcommand{\natural}{{\vrule height 1.6ex width
.05em depth 0ex \kern -.35em {\rm N}}}

\newcommand{\tr}{{\rm {Tr} \,}}

\newcommand{\dfrac}[2]{{\displaystyle{\frac{#1}{#2}}}}
\newcommand{\dsum}[2]{\displaystyle{\sum_{#1}^{#2}}}

\newcommand{\ie}{{{\em i.e.}\ }}

\newcommand{\sla}{\raise.15ex\hbox{$/$}\kern -.57em}

\newcommand{\twiddle}{\lower.9ex\rlap{$\kern -.1em\scriptstyle\sim$}}

\newcommand{\vf}{{\varphi}}

\newcommand{\equ}[1]{(\ref{#1})}
\newcommand{\eq}{\begin{equation}}
\newcommand{\eqn}[1]{\label{#1}\end{equation}}
\newcommand{\eea}{\end{eqnarray}}
\newcommand{\eqa}{\begin{eqnarray}}
\newcommand{\eqan}[1]{\label{#1}\end{eqnarray}}
\newcommand{\ba}{\begin{array}}
\newcommand{\ea}{\end{array}}
\newcommand{\eqac}{\begin{equation}\begin{array}{rcl}}
\newcommand{\eqacn}[1]{\end{array}\label{#1}\end{equation}}

\newcommand{\at}{{\~a}}

\newcommand{\ooo}{{\'o}}

\newcommand{\iii}{{\'\i}}

\begin{document}
%********************************************************
%{\large     %POUR PREPRINT
%********************************************************
%*********************************************************************
%\def\ftoday{{\sl  \number\day \space\ifcase\month
%\or Janvier\or F\'evrier\or Mars\or avril\or Mai
%\or Juin\or Juillet\or Ao\^ut\or Septembre\or Octobre
%\or Novembre \or D\'ecembre\fi
%\space  \number\year}}
%grav.tex \hspace{1cm} {\bf DRAFT} \hfill{\ftoday}\\

%\hfill{\number\time}\\

%********************************************************
%********************************************************
%PAGE DE TITRE
%*********************************************************
{\ }

\vspace{3cm}
%***************************************************************
\centerline{\LARGE BRS Cohomology of Zero Curvature Systems}\vspace{2mm}
\centerline{\LARGE II. The Noncomplete Ladder Case}

%***************************************************************
\vspace{1cm}

\centerline{\bf {\large M. Carvalho, L.C.Q. Vilar, C.A.G. Sasaki}}
\vspace{2mm}
\centerline{\it C.B.P.F}
\centerline{\it Centro Brasileiro de Pesquisas F{\iii}sicas,}
\centerline{\it Rua Xavier Sigaud 150, 22290-180 Urca}
\centerline{\it Rio de Janeiro, Brazil}
\vspace{3mm}
\centerline{ and }
\vspace{3mm}
\centerline{\bf {\large S.P. Sorella}}
\vspace{2mm}
\centerline{{\it UERJ}}
\centerline{{\it Departamento de F{\iii}sica Te{\ooo}rica}}
\centerline{{\it Instituto de F{\iii}sica, UERJ}}
\centerline{{\it Rua S{\at}o Francisco Xavier, 528}}
\centerline{{\it 20550-013, Rio de Janeiro, Brazil}}
\centerline{ and }
\centerline{\it C.B.P.F}
\centerline{\it Centro Brasileiro de Pesquisas F{\iii}sicas,}
\centerline{\it Rua Xavier Sigaud 150, 22290-180 Urca}
\centerline{\it Rio de Janeiro, Brazil}
\vspace{10mm}

%\centerline{{\normalsize {\bf PACS: 11.15.Bt }} }
%\vspace{4mm}

\centerline{{\normalsize {\bf REF. CBPF-NF-063/95}} }

\vspace{4mm}
\vspace{10mm}

\centerline{\Large{\bf Abstract}}\vspace{2mm}
\noindent
The Yang-Mills type theories and their BRS cohomology are analysed within
the zero curvature formalism.

\setcounter{page}{0}
\thispagestyle{empty}

\vfill
\pagebreak
%\null
%%%%%%%%%%%%%%%%%%%%%%%%%%%%%%%%%%%%%%%%%%%%%%%%%%%%%%%%%%%%%%%%%%%%%%%%%%%
%%%%%%%%%%%%%%%%%%%%%%%%%%%%%%%%%%%%%%%%%%%%%%%%%%%%%%%%%%%%%%%%%%%%%%%%%%%
%%%%%%%%%%%%%%%%%%%%%%%%%%%%%%%%%%%%%%%%%%%%%%%%%%%%%%%%%%%%%%%%%%%%%%%%%%%
%%%%%%%%%%%%%%%%%%%%%%%%%%%%%%%%%%%%%%%%%%%%%%%%%%%%%%%%%%%%%%%%%%%%%%%%%%%
%%%%%%%%%%%%%%%%%%%%%
%INTRODUCAO
%%%%%%%%%%%%%%%%%%%%%
\section{Introduction}
In the first part (I) of this work we have studied the zero
curvature formulation of systems described by means of
a complete ladder field, the components of which
span all possible form degrees. The present paper is devoted to
analyse the zero curvature equation in the case in which the
completeness condition for the generalized ladder field is relaxed.
This means that we shall deal with a gauge ladder ${\wti \AA}$
for which the form degree of the highest component is strictly
lower than the space-time dimension $D$, \ie
\eq
{\wti \AA} = c + A + \vf^{-1}_2 + ...+ \vf^{1-q}_q \ ,
\qquad 1\leq q < D \ .
\eqn{ladder-cA}
As we shall see in the following, the noncomplete case will
display a set of remarkable features which will make it quite
different from the previous complete case.
The first interesting aspect, as already mentioned in the
introduction of part I, is that the consistency of the zero curvature
condition
\eq{\wti \FF} = \dT {\wti \AA}  -i  {\wti \AA}^2 =0
\eqn{zero-curv-cond}
implies now the existence of a set of new operators
$(\GG^{1-k}_k , \,\, 2\leq k\leq D )$ which are in involution,
according to the algebra
\eq\ba{l}
\GG^{-1}_2 \, = \dfrac{1}{2}[\,{\d}\,,\, d\,] \ ,\, \\[3mm]
\GG^{1-k}_k = \dfrac{1}{k}\left [\,\d \,,\, \GG^{2-k}_{k-1}\,\right ]
\ , \qquad k > 2 \ ,
\ea\eqn{invol-oper}
$\d $ being the operator which together with the BRS operator
$b$ decomposes the exterior space-time derivative $d$ as
\eq
              d=-\lc b,\d \rc \ .
\eqn{delt-def}
The second interesting feature of the noncomplete
case is that the cohomology of the BRS operator $b$ is richer
than the corresponding one of the complete case. \linebreak
\noindent
Indeed, the
noncompleteness of ${\wti \AA}$ will allow us to introduce a set
of curvatures $(R^{1-m}_{m+1} , \,\, 1\leq m\leq q )$ which are
a generalization of the familiar two-form gauge field strength
$F = d A  -i A^2$. It follows then
that, in addition to the usual ghost cocycles $(\tr c^{2n+1})$ of
the complete case (see Sect. 4 of I), the cohomology of $b$ now includes
also invariant polynomials in the highest curvature $(R^{1-q}_{q+1})$.

As a consequence of these new features, the expressions of the
polynomials \linebreak $\o^{G+D-j}_j  \,\, (0\leq j\leq D )$ which
solve the descent equations
\eq\ba{l}
    b\, \o^{G+j}_{D-j} + d\, \o^{G+j+1}_{D-j-1}  = 0 \ ,
    \qquad 0\leq j \leq (D-1) \ ,\es
    b\, \o^{G+D}_{0}                       = 0 \ ,
\ea\eqn{desc-eq}
will get modified with respect to the complete case. This
modification will result in the appearence of a set of local
polynomials $\O^{G+D-j}_j  \,\, (q+1\leq j\leq D )$ in the
curvatures $(R^{1-m}_{m+1})$ which have to be added to the
cocycles obtained from the expansion of the generalized terms
$(\tr {\wti \AA}^{G+D})$.
These polynomiYang-Mills, turn out to be
characterized by a set of consistency conditions involving the
operators $\GG^{1-k}_k$.

The second part of the work is organized as follows. In Sect. 2
we present the zero curvature condition for the noncomplete gauge
ladder. Sect. 3 is devoted to the study of the cohomology of the
BRS operator. In Sect. 4 we solve the descent equations. Sect. 5 and
Sect. 6 are finally devoted to the discussion of several examples
among which one finds the zero curvature formulation of the pure
Yang-Mills gauge theory.

%%%%%%%%%%%%%%%%%%%%%%%%%%%%%
%ZERO CURVATURE
%%%%%%%%%%%%%%%%%%%%%%%%%%%%%

\section{The zero curvature condition}

In part I (cf. Sect. 2) the BRS transformations of the various
components of the gauge ladder ${\wti \AA}$ have been obtained
by constraining the latter to obey a zero curvature condition.
Equivalently, as we have seen in Sect. 3 of I, once the BRS
transformations of the fields have been given, the zero curvature condition
becomes a consequence of the existence of the operator $\d$
which realizes the decomposition \equ{delt-def}.
This second procedure will be taken as the starting point for
the discussion of the zero curvature condition in the present
noncomplete case. The gauge ladder ${\wti \AA}$ takes now the
following form
\eq
{\wti \AA} = c + A + \vf^{-1}_2 + ...+ \vf^{1-q}_q \ ,
\qquad 1\leq q < D \ ,
\eqn{ladder-cA2}
$D$ being the dimension of the space-time.
We will assume therefore that the nilpotent BRS transformations
of the components $\vf^{1-j}_j  \,\, (0 \leq j \leq q )$ of
\equ{ladder-cA2} will be the same as those of the corresponding complete case
(see Sect. 2 of I), \ie
\eq\ba{l}
  b  c= ic^2 \ , \es
  b  A= -d c + i [c,A] \ , \es
  b \vf^{1-j}_j =-d \vf^{2-j}_{j-1} + \dfrac{i}{2} \dsum{m=0}{j}
    \lc \vf^{1-m}_m\,,\,\vf^{1-j+m}_{j-m} \rc  \ ,
  \qquad   2 \leq j \leq q \ ,
\ea\eqn{brs-transform-cAladder}
where, as usual, $\lc a , b \rc = ab-(-1)^{|a||b|}ba$ denotes the graded
commutator and,  as done in I, we shall work in the functional space
$\VV$ of form-valued polynomials built up with the fields $\vf^{1-j}_j$ and
their differentials $d\vf^{1-j}_j$, \ie
\eq
\VV = \hbox{polynomials in}\,\,
\lp \vf^{1-j}_j,\,d\,\vf^{1-j}_j \,;\,\,\, 0\leq j \leq q \rp \ .
\eqn{base}
Having assigned the BRS transformations, let us turn to the introduction of
the decomposition \equ{delt-def}. To this purpose we
define the operator $\d$ as
\eq\ba{l}
{\wti \AA} = e^{\d}c \ , \es
    \d \, \vf^{1-j}_j = (j+1)\vf^{-j}_{j+1}\ , \qquad 0\leq j \leq q-1 \ , \es
   \d \vf^{1-q}_q = 0 \ ,
\ea\eqn{field-delta-action}
and
\eq\ba{l}
    \d \,d\, \vf^{1-m}_m = (m+1)\,d\,\vf^{-m}_{m+1} \ ,
    \qquad 0\leq m \leq q-2 \ , \es
    \d \,d\, \vf^{2-q}_{q-1} = q\,d\,\vf^{1-q}_q -
    (q+1)\lp d\,\vf^{1-q}_q - \dfrac{i}{2}\dsum{j=1}{q}
    \lc \vf^{1-j}_j , \vf^{j-q}_{q-j+1} \rc \rp \ , \es
    \d \,d\, \vf^{1-q}_{q} =\dfrac{i}{2}(q+1)\dsum{j=1}{q}
     \lc \vf^{-1-q+j}_{q+2-j} , \vf^{1-j}_j \rc \ .
\ea\eqn{diff-field-delta-action}

One easily checks that, on the functional space $\VV$, the
operators $b$ and $\d$ realize the decomposition
\equ{delt-def}, \ie
\eq
              d=-\lc b,\d \rc \ .
\eqn{delt-def3}
Comparing now equations \equ{field-delta-action},
\equ{diff-field-delta-action} with the corresponding ones of the
complete ladder case (see Sect. 2 of I) one sees that, while the
action of the operator $\d$ on the components $(\vf^{1-j}_j)$ is
the same, the transformations of the differentials of higher
form-degree, \ie $(d\vf^{2-q}_{q-1})$ and $(d\vf^{1-q}_{q})$,
are now nonvanishing.
This fact implies that, contrary to the complete case, the
operator $\d$ does not commute anymore with the exterior derivative $d$,
\eq
\lc \d , d \rc \neq 0 \ .
\eqn{delta-deconp-not}
In addition, depending on the dimension of the space-time $D$
and on the number $q$ of components of the gauge ladder
${\wti\AA}$, the commutators
\eq
\lc \d , \lc \d ,\lc \d ,.....,d\rc \rc \rc
\eqn{commutator1}
turn out to be nonvanishing as well.

This algebraic structure, which generalizes that
of ref.~\cite{dec,tat}, will have important consequences on the zero
curvature condition. The latter, repeating the same argument of
Sect. 3 of I, is obtained by applying the operator $e^{\d}$ on
the BRS transformation of the zero-form ghost field $c$, \ie
\eq
 e^{\d}\,b\,e^{-\d} e^{\d}c =   e^{\d} \,i \, c^2 \ .
\eqn{delta-transform-of-c}
Recalling now that ${\wti \AA}=e^{\d}c$ and defining the generalized
operator $\dT$ as
\eq
\dT = e^{\d}\,b\,e^{-\d} \ ,
\eqn{dtild}
we get the zero curvature condition
\eq
\dT {\wti \AA}  = i  {\wti \AA}^2
\eqn{zero-curv-condition}
for the noncomplete ladder case. Equation \equ{zero-curv-condition}
is, however, only apparently similar to the corresponding
condition of the complete ladder case. In fact, due to
eqs.\equ{delta-deconp-not} and \equ{commutator1}, the operator
$\dT$ is now given by
\eq
\dT = b + d + \dsum{n\geq 2}{D}\,\dfrac{1}{n!}
\underbrace{\lc \d , \lc \d ,\lc \d ,.....,d\rc \rc
\rc}_{\hbox{(n-1)-times}} \ ,
\eqn{dtidef}
so that, defining the operators
\eq\ba{l}
\GG^{-1}_2 \, = \dfrac{1}{2}[\,{\d}\,,\, d\,] \ ,\, \\[3mm]
\GG^{-2}_3 = \dfrac{1}{3!}\lc\,\d \lc \d , d \rc \rc =
\dfrac{1}{3}\lc \d , \GG^{-1}_2 \rc \ ,\, \\[3mm]
\GG^{-3}_4 = \dfrac{1}{4!}\lc \d ,\lc \d , \lc \d , d \rc \rc \rc =
\dfrac{1}{4}\lc \d , \GG^{-2}_3 \rc \ , \\[3mm]
\qquad  ..................
\ea\eqn{invol-oper0}
we have
\eq
\dT = b + d + \dsum{k\geq 2}{D}\GG^{1-k}_k
\eqn{dtild-gg}
with
\eq\ba{l}
\GG^{-1}_2 \, = \dfrac{1}{2}[\,{\d}\,,\, d\,] \ ,\, \\[3mm]
\GG^{1-k}_k = \dfrac{1}{k}\left [\,\d \,,\, \GG^{2-k}_{k-1}\,\right ]
\ , \qquad k > 2 \ .
\ea\eqn{invol-operator}

One thus sees that in the noncomplete case the
zero curvature condition is accompanied by a set of operators
$\GG^{1-k}_k$ which are in involution, according to eq.\equ{invol-operator}.
We underline, in particular, that the origin of the operators $\GG^{1-k}_k$
actually relies on the noncomplete character of the gauge ladder
\equ{ladder-cA2}. It is very easy, using the equations
\equ{field-delta-action} and \equ{diff-field-delta-action}, to derive the
explicit form of the various operators  $\GG^{1-k}_k$ appearing in the
eq. \equ{zero-curv-condition}. In particular, as we shall show later on in the
examples, the number of operators  $\GG^{1-k}_k$ which do not identically
vanish depends both on the dimension $D$ of the space-time and on the
number $q$ of components of the gauge ladder ${\wti \AA}$.
We also notice that these operators are absent when $q=D$, \ie they are not
present in the case in which
the ladder is complete.

Moreover their existence implies that the cohomology of the
operator $\dT$ is no more directly related to that of the
operator $(d+b)$.
Therefore the cohomology classes of $\dT$ do not immediately
provide solutions of the descent equations \equ{desc-eq}.
It turns out indeed that in order to obtain a solution of the
tower \equ{desc-eq} we must add to the cohomology classes of
$\dT$, \ie $\tr {\wti \AA}^{2n+1}$,
certain polynomials
$\O^{G+D-j}_j  \,\, (q+1\leq j\leq D)$ which obey a set of consistency
conditions involving the
operator $\GG^{1-k}_k$. In other words, the presence of the
$\GG^{1-k}_k$'s requires a modification of the solution of the
descent equations with respect to the complete ladder case (see
Sect. 5 of I).

Let us conclude this section with the following remark. Instead
of having assumed the BRS transformations \equ{brs-transform-cAladder}
we could have started directly with the zero curvature condition
\equ{zero-curv-condition}.
It is easily verified then that the introduction of the
operators $\GG^{1-k}_k$ is needed in order to avoid the
appearence of constraints among the components of the
noncomplete ladder field ${\wti \AA}$.

%%%%%%%%%%%%%%%%%%%%%
%BRS COHOMOLOGY
%%%%%%%%%%%%%%%%%%%%%
\section{Cohomology of the BRS operator}

The first step in order to solve the descent equations \equ{desc-eq}
is that of computing the cohomology of the BRS operator $b$.
This task, due to the noncomplete character of ${\wti \AA}$,
will turn out to be simplified by the introduction of the
following curvatures $R^{1-m}_{m+1}$ of total degree two:
\eq
R^{1-m}_{m+1} = d\,\vf^{1-m}_{m} - \dfrac{i}{2}\dsum{k=1}{m}
\lc \vf^{1-k}_{k} , \vf^{k-m}_{m+1-k}\rc ; \qquad 1\leq m \leq q
\ .
\eqn{curv-def}
In particular, for $m=1$ the expression \equ{curv-def} reduces to
\eq
R^0_2 = d A - i A^2 = F \ ,
\eqn{z}
\ie one recovers the familiar two-form gauge field strength. We
also remark that, for $m > 1$, the curvatures $R^{1-m}_{m+1}$
possess the property of having  negative ghost number.

The great advantage of working with the curvatures
$R^{1-m}_{m+1}$ relies on the fact that they transform
covariantly under the action of the BRS operator, \ie
\eq
b\,R^{1-m}_{m+1}= i \lc c , R^{1-m}_{m+1}\rc \ .
\eqn{covar-trans-curv}
This feature, following the well known
Yang-Mills case~\cite{russian1,bdk,dvhtv,bbh}, suggests that it is
convenient to use the curvatures
$R^{1-m}_{m+1}$ as independent variables instead of the
differentials $d\vf^{1-m}_{m}$, \ie we replace everywhere the
variables $d\vf^{1-m}_{m}$ by $R^{1-m}_{m+1}$ making use of
eq.\equ{curv-def}.
Consequently, for the functional space $\VV$ we have
\eq
\VV = \hbox{polynomials in}\,\,
 \Lp \,  c \ ,  A \ , \vf^{1-m}_m,\,\,\, 2\leq m \leq q \ ; dc \ ,
  R^{1-j}_{j+1}, \, \, \,  1\leq j \leq q \, \Rp  \ ,
\eqn{base2}
and, for the nilpotent BRS transformations,
\eq\ba{l}
  b\,c= ic^2 \ , \es
  b\,A= -d c + i [c,A] \ , \es
  b\,\vf^{1-m}_m = i \lc c , \vf^{1-m}_m \rc - R^{2-m}_m \ ,
  \qquad 2 \leq m \leq q \ , \es
  b\,d\,c = i \lc c , dc \rc \ , \es
  b\,R^{1-j}_{j+1} = i \lc c , R^{1-j}_{j+1} \rc \ , \qquad
1\leq j \leq q \ .
\ea\eqn{brs-transform-cAladder5}
Let us turn now to the computation of the cohomology of $b$.
Introducing the filtering operator $\NN$
\eq\ba{l}
       \NN c = c \ , \qquad  \NN A = A \ ,  \es
      \NN \vf^{1-m}_m = \vf^{1-m}_m
       \ , \qquad 2 \leq m \leq q \ , \es
       \NN \,d\,c = dc \ , \qquad  \NN R^{1-j}_{j+1}=R^{1-j}_{j+1}\ ,
       \qquad 1 \leq j \leq q \ ,
\ea\eqn{filtering}
the BRS operator decomposes as
\eq
b = b_0 + b_1 \ ,
\eqn{decb1b0}
with
\eq\ba{l}
  b_0\,c= 0\ , \es
  b_0\,A= -d c  \ , \qquad b_0\,d\,c =0 \ ,\es
  b_0\,\vf^{1-m}_m = - R^{2-m}_m \ , \qquad b_0R^{2-m}_m =0\ ,
  \qquad 2 \leq m \leq q \ , \es
  b_0\,R^{1-q}_{q+1} = 0\ , \es
  b_0^2 = 0 \ .
\ea\eqn{b-zerobrs-transform}
Equations \equ{b-zerobrs-transform} show that all the variables
except the zero-form ghost $c$ and the highest curvature
$R^{1-q}_{q+1}$ are grouped in BRS doublets. This implies that
the cohomology of $b_0$ and, in turn, that of the full BRS
operator $b$ depend only on $c$ and $R^{1-q}_{q+1}$.
More precisely, using the general results of
refs.~\cite{russian1,bdk,dvhtv,bbh},
it follows that the cohomology of $b$ on the functional space $\VV$ is
spanned by invariant polynomials in the variables $(c,
R^{1-q}_{q+1})$ built up with factorized monomials of the form
\eq
\LP \tr \frac{c^{2n+1}}{(2n+1)!} \RP \cdot
\LP \tr (R^{1-q}_{q+1})^m \RP
\ , \qquad n,m = 1,2,....\ .
\eqn{b-cohomology}
One sees  thatof the BRS operator $b$, in addition of the usual ghost cocycles
$(\tr c^{2n+1})$, includes  also polynomials in the highest
curvatures $R^{1-q}_{q+1}$. Notice finally that, being the ghost
number of the highest curvature $R^{1-q}_{q+1}$ negative for
$q> 1$, the cohomology classes of $b$ are nonvanishing in the
negative charged sectors.

We conclude this section by remarking that the highest curvature
$R^{1-q}_{q+1}$ is actually related to the ghost field $c$ through the action
of the operator $\GG^{-q}_{q+1}$,
\eq
  \GG^{-q}_{q+1} \, c = \hbox{(const)} \, R^{1-q}_{q+1}  \ ,
\eqn{highest-curv-ghost}
the proportionality factor being easily computed by means of the
eqs.\equ{field-delta-action}, \equ{diff-field-delta-action}.

%*********************************************************************
%*********************************************************************
%*********************************************************************
%*********************************************************************

\section{Solution of the descent equations}

Having characterized the cohomology of the BRS operator $b$, let
us focus  on the cohomology of $b$ modulo $d$, \ie let us try
to solve the descent equations
\eq\ba{l}
    b\, \o^{G+j}_{D-j} + d\, \o^{G+j+1}_{D-j-1}  = 0 \ ,
    \qquad 0\leq j \leq (D-1) \ ,\es
    b\, \o^{G+D}_{0}                       = 0 \ .
\ea\eqn{desc-eq3}
As mentioned before and as already observed in the case of pure
Yang-Mills (\ie $q=1$), the presence of the operators
$\GG^{1-k}_k$ in the zero curvature condition \equ{zero-curv-condition}
requires a slight modification of the climbing procedure
presented in the previous  complete ladder case (see I).

Repeating indeed the same argument of~\cite{dec}, it is easy to
convince oneself that, once a solution $\o_0^{G+D}$ of thequation of
\equ{desc-eq3} has been obtained, an explicit
expression for the higher polynomials $\o_j^{G+D-j}$ is provided by
the generalized cocycle ${\tilde \o}^{G+D}$ of total degree $(G+D)$
\eq\ba{l}
{\tilde \o}^{G+D} = \dsum{j=0}{D} \, \o^{G+D-j}_j \ , \es
{\tilde \o}^{G+D} = e^{\d }\lp {\o}^{G+D}_0 +
\dsum{j=q+1}{D} \O^{G+D-j}_j \rp \ ,
\ea\eqn{omegatilde}
where ${\o}^{G+D}_0$ is
\eq
{\o}^{G+D}_0 = \tr \dfrac{c^{G+D}}{(G+D)!} \ ,
\eqn{omegazerogd}
and the quantities $\O^{G+D-j}_j$ are determined recursively by means of
the consistency conditions
\eq\left \{\ba{l}
b\,\O^{G+D-j}_j = (j-1)(-1)^j\, \GG^{1-j}_j\o^{G+D}_0 +
\dsum{k=2}{(j-1)} (k-1)(-1)^k \GG^{1-k}_k \O^{G+D-j+k}_{j-k}\ , \es
\O^{G+D-j+k}_{j-k} = 0 \qquad \hbox{if} \qquad (j-k)< q+1 \ .
\ea\right .\eqn{Omgr}
As we shall see, the latters turn out to be easily disentangled by using the
results \equ{b-cohomology} on the BRS cohomology.
Moreover, setting $q=1$, equations \equ{Omgr} are seen to
reproduce those already met in the pure Yang-Mills case~\cite{dec}.
In particular, from equations \equ{omegatilde} and \equ{omegazerogd},
we see that the solution of the tower \equ{desc-eq3} in the
noncomplete case turn out to be deformed  with respect to the corresponding
solution of the complete ladder case (see Sect. 5 of I) by the
inclusion of the cocycles $\O^{G+D-j}_j$.

%Let us finally  remark that, depending on the
%dimension of the space-time and on the ghost number of
%the highest curvature $R^{1-q}_{q+1}$, the descent equations
%\equ{desc-eq3} may display an obstruction for higher negative
%values of $G$. In this case, as it is well known~\ref{}, the
%general solution of \equ{desc-eq3} does not receive
%contributions coming from the lowest level.

\section{Example I: Pure Yang-Mills theory as a zero curvature system}

As a first important example of a noncomplete ladder system let
us present here the zero curvature formulation of the pure Yang-Mills
gauge theory in any space-time dimension, corresponding to a
generalized ladder with $q=1$, \ie
\eq
{\wti \AA} = c + A \ .
\eqn{ncladderYM}
It is worthy to recall that, since the Yang-Mills theories are
power-counting nonrenormalizable for space-time dimensions greater than
four, the fields $A$ and $c$, unlike the three dimensional Chern-Simons case
discussed in I, are now regarded as unquantized external fields coupled to
currents of quantum matter fields. Therefore, the existence of gauge
anomalies at the quantum level, will correspond to a violation of the
conservation law of the matter currents and to the appearence of Schwinger
terms in the corresponding current algebra.

It is easily checked that in this case the consistency of the zero
curvature condition \equ{zero-curv-condition} requires that only the
first operator
$\GG^{-1}_2$ of eq.\equ{dtild-gg} is nonvanishing.
Therefore for the operator $\dT$ we get
\eq
\dT = b + d + \GG^{-1}_2 \ ,
\eqn{dddtttYM}
and from
\eq
    \dT {\wti \AA}\, = \, i\, {\wti \AA}^2  \ ,
\eqn{zero-curvature-again}
we obtain
\eq\ba{l}
  b\,c= ic^2 \ , \es
  b\,A= - d\,c + i [c,A] \ ,
\ea\eqn{brs-transform-cAladder10}
and
\eq\ba{l}
\GG^{-1}_2 \,c  =  -d\,A + i\,A^2 = - F \ , \es
\GG^{-1}_2 \,d\,c = i [ A , F ] \ , \es
\GG^{-1}_2\, A =  \GG^{-1}_2 \,F = 0\ .
\ea\eqn{invol-oper2}
{}From equations  \equ{field-delta-action}, \equ{diff-field-delta-action},
for the operator $\d$ we have
\eq\ba{l}
    \d \, c = A \ , \qquad \d \, dc = -d\,A + 2iA^2 \ , \es
    \d A =0 \ , \qquad \d \,d A = 0 \ ,
\ea\eqn{diff-field-delta-action-YM}
and
\eq\ba{l}
    d = - [ d , \d ] \ , \qquad \GG^{-1}_2 = \dfrac{1}{2}[ \d , d ] \ , \\[3mm]
    [ \d , \GG^{-1}_2 ] = \, [ b , \GG^{-1}_2 ] = \,  [ d , \GG^{-1}_2] = 0 \ .
\ea\eqn{dgdelta-algebra}
For what concerns the solutions of the descent equations \equ{desc-eq3}
here we shall limit ourselves only to state the final result, reminding
the reader to the detailed discussion and proofs already given in~\cite{dec}.
We underline in particular that, as proven in~\cite{tat}, the cocycles
$\O^{G+D-j}_j$ appearing in eq.\equ{omegatilde} can be summed up
into a unique closed
generalized expression which collects both the gauge anomalies
and the Chern-Simons terms. The latters are given respectively by
\eq
\o^1_{2n}=\dsum{p=0}{n}\dfrac{i^{(n-p)}}{(2n-p+1)!p!} \LP \PP \lp
c, F^p, (A^2)^{n-p}\rp + i(n-p)\PP \lp [c , A ], F^p, A, (A^2)^{n-p}\rp
\RP \ ,
\eqn{cs-anomalies-par}
and
\eq
\o^0_{2n+1}=\dsum{p=0}{n}\dfrac{i^{(n-p)}}{(2n-p+1)!p!} \, \PP \lp
 F^p, A, (A^2)^{n-p} \rp \ ,
\eqn{cs-anomalies-impar}
where the integer $n=1,2,...$ labels the various dimensions of the
space-time and $\PP ({\cal \JJ}_1, {\cal \JJ}_2,..., {\cal \JJ}_n)$ denotes
the symmetric invariant polynomials defined as
\eq
\PP ({\cal \JJ}_1, {\cal \JJ}_2,..., {\cal \JJ}_n) =
{\cal \JJ}_1^{a_1}{\cal \JJ}_2^{a_2}......{\cal \JJ}_n^{a_n} \,
S\tr ( T^{a_1}T^{a_2}.....T^{a_n})\ ,
\eqn{sym-inv-poly}
$S\tr $ being the symmetrized trace~\cite{russian} and, following
Zumino's notations~\cite{zum}, we have used
\eq
\PP ({\cal \JJ}_1, {\cal \JJ}_2,..., {\cal \JJ}^p) =
\PP ({\cal \JJ}_1, {\cal \JJ}_2, {\cal \JJ}_3, \underbrace{{\cal \JJ},...,
{\cal \JJ}}_{\hbox{p-times}}) \ .
\eqn{sym-pol}
It is worthy to emphasize that, actually, the formulas
\equ{cs-anomalies-par}, \equ{cs-anomalies-impar} represent one
of the most compact expression for the gauge
anomaly and for the Chern-Simons term in any space-time dimension.

%**************************************************
%**************************************************
%****************************
\section{Example II: the case $D=6$, $G=1$, $q=3$}

In order to clarify the role of the operators $\GG^{1-k}_k$ and
of the generalized curvatures $R^{1-m}_{m+1}$, let us discuss in
this second example the solution of the descent equations \equ{desc-eq}
in the six dimensional case $D=6$ with ghost number $G=1$ and a
gauge ladder with $q=3$, \ie
\eq
{\wti \AA} = c + A + \vf^{-1}_2 + \vf^{-2}_3 \ .
\eqn{ladder}
{}From eqs.\equ{brs-transform-cAladder5}, for the BRS
transformations we have
\eq\ba{l}
  b\,c= ic^2 \ , \es
  b\,A= -d c + i [c,A] \ , \es
  b\,\vf^{-1}_2 = i \lc c , \vf^{-1}_2 \rc - R^0_2 \ , \es
  b\,\vf^{-2}_3 = i \lc c , \vf^{-2}_3 \rc - R^{-1}_3 \ ,
\ea\eqn{brs-transform-example2}
where $R^0_2$, $R^{-1}_3$ are the generalized curvatures of
eq.\equ{curv-def} whose expressions are given
\eq\ba{l}
R^0_2 = F = dA - i A^2 \ , \es
R^{-1}_3 = d\vf^{-1}_2 - i \lc A , \vf^{-1}_2 \rc \ .
\ea\eqn{curv-example2}
In particular, for the highest curvature $R^{-2}_4$ we have
\eq
R^{-2}_4 = d\vf^{-2}_3 - i \lc A , \vf^{-2}_3 \rc -
\dfrac{i}{2} \lc \vf^{-1}_2 , \vf^{-1}_2 \rc
\eqn{curv-max}
and
\eq
b\, R^{1-m}_{m+1} = i \lc c , R^{1-m}_{m+1} \rc \ ,
\qquad 1 \leq m \leq 3 \ .
\eqn{curv-gen}
The curvatures $(R^{0}_2, R^{-1}_3, R^{-2}_4)$ obey the
following generalized Bianchi identities
\eq\ba{l}
d\,R^{0}_2 = i \lc A , R^{0}_2 \rc \es
d\,R^{-1}_3 = i \lc A , R^{-1}_3 \rc +
i \lc \vf^{-1}_2 , R^{0}_2 \rc \es
d\,R^{-2}_4 = i \lc A , R^{-2}_4 \rc +
i \lc \vf^{-1}_2 , R^{-1}_3 \rc +
i \lc \vf^{-2}_3 , R^{0}_2 \rc \ .
\ea\eqn{dcurv-example}
They transform under the operator $\d$ of eqs.
\equ{field-delta-action}, \equ{diff-field-delta-action} as
\eq\ba{l}
\d\,R^{0}_2 = 2 R^{-1}_3 \es
\d\,R^{-1}_3 = - R^{-2}_4 - \dfrac{i}{2} \lc \vf^{-1}_2 , \vf^{-1}_2 \rc \es
\d\,R^{-2}_4 = -i \lc \vf^{-2}_3 , \vf^{-1}_2 \rc \ .
\ea\eqn{delta-curv-example}
For what concerns the operators $\GG^{1-k}_k$ of eq.\equ{invol-operator}
it is easily seen that in the present example the zero curvature
equation \equ{zero-curv-condition} implies the existence of a
set of five nonvanishing operators
$(\GG^{-1}_2, \GG^{-2}_3, \GG^{-3}_4, \GG^{-4}_5, \GG^{-5}_6)$.
Their action on the fields and on the curvatures is given
respectively by
\eq \left \{ \ba{l}
\GG^{-1}_2\,c = 0 \ , \qquad \GG^{-1}_2\,A = 0 \ ,
    \qquad \GG^{-1}_2\,\vf^{-1}_2
= -2R^{-2}_4  \ , \es
\GG^{-1}_2\,\vf^{-2}_3 = 2i \lc \vf^{-2}_3 , \vf^{-1}_2 \rc  \ ,  \es
\GG^{-1}_2\,dc = 0 \ , \qquad \GG^{-1}_2\,R^{0}_2 = 0  \ , \es
\GG^{-1}_2\,R^{-1}_3 = 2i  \Lp \lc \vf^{-1}_2 , R^{-1}_3 \rc +
\lc \vf^{-2}_3 , R^{0}_2 \rc \Rp   \ , \\[3mm]
\GG^{-1}_2\,R^{-2}_4 = 2i \lc \vf^{-2}_3 , R^{-1}_3 \rc \ ,
\ea \right .\eqn{g2-action}

\eq \left \{ \ba{l}
\GG^{-2}_3\,c = 0 \ ,  \qquad \GG^{-2}_3\,A = \dfrac{4}{3}R^{-2}_4 \es
\GG^{-2}_3\,\vf^{-1}_2 = -\dfrac{4i}{3} \lc \vf^{-2}_3 ,
\vf^{-1}_2 \rc \ ,  \es
\GG^{-2}_3\,\vf^{-2}_3 = 2i \lc \vf^{-2}_3 , \vf^{-2}_3 \rc \ , \es
\GG^{-2}_3\,dc = 0  \ ,   \es
\GG^{-2}_3\,R^{0}_2 = -\dfrac{4i}{3} \Lp \lc \vf^{-1}_2 , R^{-1}_3 \rc +
\lc \vf^{-2}_3 , R^{0}_2 \rc \Rp \ ,   \\[3mm]
\GG^{-2}_3\,R^{-1}_3 = 4i \lc \vf^{-2}_3 , R^{-1}_3 \rc  \ , \es
\GG^{-2}_3\,R^{-2}_4 = 0 \ ,
\ea \right .\eqn{g3-action}

\eq \left \{ \ba{l}
\GG^{-3}_4\,c = -\dfrac{1}{3}R^{-2}_4 \ ,  \qquad \GG^{-3}_4\,A = \dfrac{i}{3}
\lc \vf^{-2}_3 , \vf^{-1}_2 \rc  \ , \\[3mm]
\GG^{-3}_4\,\vf^{-1}_2 = -\dfrac{5i}{2} \lc \vf^{-2}_3 \ ,
\vf^{-2}_3 \rc\ , \qquad \GG^{-3}_4\,\vf^{-2}_3 = 0 \ , \es
\GG^{-3}_4\,dc = \dfrac{i}{3} \Lp \lc A , R^{-2}_4 \rc + \lc
\vf^{-1}_2 , R^{-1}_{3} \rc + \lc \vf^{-2}_3 , R^{0}_2 \rc \Rp \ ,  \\[3mm]
\GG^{-3}_4\,R^{0}_2 = \dfrac{i}{3} \Lp  \lc \vf^{-1}_2 , R^{-2}_4
\rc - 11 \lc \vf^{-2}_3 , R^{-1}_3 \rc \Rp  \ , \\[3mm]
\GG^{-3}_4\,R^{-1}_3 = 0 \ ,  \qquad \GG^{-3}_4\,R^{-2}_4 = 0 \ ,
\ea \right .\eqn{g4-action}

\eq \left \{ \ba{l}
\GG^{-4}_5\,c = 0 \ ,  \qquad \GG^{-4}_5\,A = \dfrac{6i}{5}
\lc \vf^{-2}_3 , \vf^{-2}_3 \rc \ , \\[3mm]
\GG^{-4}_5\,\vf^{-1}_2 = 0 \ ,  \qquad \GG^{-4}_5\,\vf^{-2}_3 = 0  \ , \\[3mm]
\GG^{-4}_5\,dc = \dfrac{16i}{5}   \lc \vf^{-2}_3 , R^{-1}_3 \rc  \ , \\[3mm]
\GG^{-4}_5\,R^{0}_2 = \GG^{-4}_5\,R^{-1}_3 = \GG^{-4}_5\,R^{-2}_4 = 0 \ ,
\ea \right .\eqn{g5-action}

\eq \left \{ \ba{l}
\GG^{-5}_6\,c = -\dfrac{i}{5} \lc \vf^{-2}_3 , \vf^{-2}_3 \rc  \ , \\[3mm]
\GG^{-5}_6\,A = \GG^{-5}_6\,\vf^{-1}_2 = \GG^{-5}_6\,\vf^{-2}_3 = 0  \ ,
\\[3mm]
\GG^{-5}_6\,dc = \GG^{-5}_6\,R^{0}_2 = \GG^{-5}_6\,R^{-1}_3 =
\GG^{-5}_6\,R^{-2}_4 = 0  \ .
\ea \right .\eqn{g6-action}

\noindent
Turning now to the descent equations
\eq\ba{l}
b\,\o^{1+j}_{6-j} + d\,\o^{2+j}_{5-j} = 0 \ , \qquad 0\leq j \leq 5 \ ,  \es
b\,\o^7_0 = 0 \ ,
\ea\eqn{b5b6}
we have that, taking into account the result \equ{covar-trans-curv} on the
cohomology of the BRS operator $b$ and the equation \equ{omegatilde}, a
solution of the ladder \equ{b5b6} is provided by the generalized cocycle
of total degree seven
\eq
{\tilde \o}^7 = e^{\d}\lp \o^7_0 + \O^3_4 + \O^2_5 + \O^1_6 \rp
\eqn{sol-exemple-II}
with
\eq
\o^7_0 = \tr \dfrac{c^7}{7!}  \ ,
\eqn{omega-7-0}
and $(\O^3_4, \, \O^2_5, \, \O^1_6)$ solutions of the equations \equ{Omgr}, \ie
\eq
   b \, \O^3_4 = 3 \, \GG^{-3}_4 \o^7_0   \ ,
\eqn{1-second-Om-tower}
\eq
   b \, \O^2_5 = -4 \, \GG^{-4}_5 \o^7_0   \ ,
\eqn{2-second-Om-tower}
\eq
   b \, \O^1_6 = \GG^{-1}_2 \O^3_4 \, + \, 5 \, \GG^{-5}_6 \o^7_0 \ .
\eqn{3-second-Om-tower}
This system can be easily solved by using the cohomology of $b$. Indeed,
beggining with the first equation \equ{1-second-Om-tower} we have from
\equ{g4-action}
\eq
    \GG^{-3}_4 \, \o^7_0 = - \dfrac{1}{6!} \tr \dfrac{R^{-2}_4 c^6}{3} \ ,
\eqn{gg34o70}
so that $\O^3_4$ may be identified with
\eq
   \O^3_4 = - \dfrac{i}{6!} \tr\, R^{-2}_4 c^5   \ .
\eqn{O34-expression}
Concerning now the second equation \equ{2-second-Om-tower}, we get from
\equ{g5-action} that
\eq
    \GG^{-4}_5 \, \o^7_0 = 0 \ .
\eqn{gg45o70}
Moreover, since the cohomology of $b$ in the sector of form degree five and
ghost number two is empty, we may choose $\O^2_5$ to be vanishing as well
\eq
\O^2_5 = 0   \ .
\eqn{O25-expression}
Finally for the last equation \equ{3-second-Om-tower}, we get
\eq
  b \, \O^1_6 = \dfrac{2}{6!} \tr
     \Lp \, \lc \vf^{-2}_3 , R^{-1}_3 \rc \, c^5 \, - \,
      i \vf^{-2}_3 \vf^{-2}_3 c^6 \, \Rp \ .
\eqn{b-O16-action}
However, from
\eq
  b\, \Lp \tr  \vf^{-2}_3 \vf^{-2}_3 c^5  \Rp =  \tr
     \Lp \, \lc \vf^{-2}_3 , R^{-1}_3 \rc \, c^5 \, - \,
      i \vf^{-2}_3 \vf^{-2}_3 c^6 \, \Rp  \ ,
\eqn{b-vf-23}
we obtain
\eq
 \O^1_6 = \dfrac{1}{6!}
          \tr \Lp \, \lc \vf^{-2}_3,  \vf^{-2}_3 \rc c^5 \, \Rp \ .
\eqn{O16-expression}
Summarizing, an explicit expression for the $\O$'s is given by
\eq\ba{l}
 \O^3_4 = - \dfrac{i}{6!} \tr\, R^{-2}_4 c^5 \ , \\[3mm]
 \O^2_5 = 0   \ ,  \\[3mm]
 \O^1_6 = \dfrac{1}{6!}
          \tr \Lp \, \lc \vf^{-2}_3,  \vf^{-2}_3 \rc c^5 \, \Rp \ .
\ea\eqn{Omega's-expression}
Of course, the above expressions are always determined modulo trivial
$b$-cocycles.

\noindent
Concluding, for the generalized cocycle ${\tilde \o}^7$  we have
\eq
{\tilde \o}^7  = \tr \Lp \,
  \dfrac{{\wti \AA}^7}{7!} \,+ \, \O^3_4 \, + \, \d \O^3_4 \, + \,
           \dfrac{\d^2}{2} \O^3_4 \, + \, \O^1_6 \, \Rp \ .
\eqn{finla-otilde-7}
The expansion of ${\tilde \o}^7$ in terms of components of different degree
and ghost number will give an explicit expression for the cocycles
entering the descent equations \equ{b5b6}.
%%%%%%%%%%%%%%%%%%%%%%%%%%%%%%%%%%%%%%%%%%%%%%%%%%%%%%%%%%%%%%%%%%%%%%%%%%
%%%%%%%%%%%%%%%%%%%%%%%%%%%%%%%%%%%%%%%%%%%%%%%%%%%%%%%%%%%%%%%%%%%%%%%%%%%
%%%%%%%%%%%%%%%%%%%%%%%%%%%%%%%%%%%%%%%%%%%%%%%%%%%%%%%%%%%%%%%%%%%%%%%%%%%
%%%%%%%%%%%%%%%%%%%%%%%%%%%%%%%%%%%%%%%%%%%%%%%%%%%%%%%%%%%%%%%%%%%%%%%%%%%
\section*{Conclusion}

We have shown that the Yang-Mills type theories can be characterized by means
of a noncomplete gauge ladder field constrained to obey a zero curvature
condition, which implies the existence of a set of new operators $\GG^{1-k}_k$.
These operators give rise together with the BRS operator $b$ to a kind of
descent equations which are easily solved using the results on the
cohomology of $b$. These solutions provide a deformation
of the cohomology of $b$ modulo $d$ with respect to the corresponding
complete ladder case presented in the part I.

%%%%%%%%%%%%%%%%%%%%%%%%%%%%%%%%%%%%%%%%%%%%%%%%%%%%%%%%%%%%%%%%%%%%%%%%%%
%%%%%%%%%%%%%%%%%%%%%%%%%%%%%%%%%%%%%%%%%%%%%%%%%%%%%%%%%%%%%%%%%%%%%%%%%%%
%%%%%%%%%%%%%%%%%%%%%%%%%%%%%%%%%%%%%%%%%%%%%%%%%%%%%%%%%%%%%%%%%%%%%%%%%%%
%%%%%%%%%%%%%%%%%%%%%%%%%%%%%%%%%%%%%%%%%%%%%%%%%%%%%%%%%%%%%%%%%%%%%%%%%%%

%}    %POUR PREPRINT
%*********************************************************************
\end{document}